\documentclass[10pt,conference,compsoc]{IEEEtran}

\usepackage[ansinew]{inputenc}
\usepackage{graphics,graphicx,color,colortbl}
\usepackage{cite}
\usepackage{url}
\usepackage{amsmath,amssymb,amsfonts,amsbsy}
\usepackage{bm}

\usepackage{float}

\newtheorem{prop}{Proposition}[section]

\newtheorem{remark}{Remark}[section]

\begin{document}

\title{Attenuating the Impact of Integrity Attacks \\ on Real-Time Pricing in Smart Grids}
\author{Jairo Giraldo and Alvaro C\'ardenas  and Nicanor Quijano}

\maketitle

\begin{abstract}

The vulnerability of false data injection attacks on real-time electricity pricing for the power grid market has been recently explored. Previous work has focused on the impact caused by attackers that compromise pricing signals and send false prices to a subset of consumers. In this paper we extend previous work by considering a more powerful and general adversary model, a new analysis method based on sensitivity functions, and by proposing several countermeasures that can mitigate the negative impact of these attacks. Countermeasures  include adding a low-pass filter to the pricing signals, selecting the time interval between price updates, selecting parameters of the controller, designing robust control algorithms, and by detecting anomalies in the behavior of the system.

\end{abstract}

\section{Introduction}

The objective of the power grid is to generate and then deliver enough electric power to match the demand of consumers. Unlike other critical infrastructures like water or gas distribution networks that can accommodate a variation in demand by storing their resource, the power grid cannot store electricity, and thus, electricity must be generated in the exact moment that it is consumed. If the supply of power is greater than the demand, this excess power is stored in the form of kinetic energy in the electricity generators, which produces an acceleration of the generator resulting in higher rotation frequency; on the other hand, if the supply of power is not enough to match the demand, generators will have to provide more current to the system, and the magnetic field associated with this increased current will slow down the generator--resulting in lower rotation frequency. All the equipment in the power grid is meant to operate at a specific frequency (e.g., 60Hz) and changes in the frequency of electricity will result in poor \emph{power quality} and ultimately risk of physical equipment damage and if the frequency deviation is large enough it may trip circuit breakers and disconnect regions of the grid causing blackouts.  


To maintain a balance between optimizing the use of resources and the real-time control requirements for keeping the frequency and voltage of the power grid at their design levels, the power grid uses a daily and hourly scheduling of generation units to match the forecast electricity load via wholesale market transactions. A scheduling coordinator solicits generation through some form of auction where lowest bidders generate electricity and this in turn creates an economically optimal schedule of generators. In contrast to these traditional wholesale markets (e.g., between generation utilities and distribution utilities), many retail markets (e.g., between a distribution utility and an industry consumer of electricity) have traditionally adopted \emph{static pricing schemes such as fixed and time-of-use tariffs, under which consumers have limited incentives to adapt their electricity consumption to market conditions. This lack of incentives results in high peak demands that strain infrastructure capacities and unnecessarily increase operational costs}~\cite{tan2013impact}. This approach is inefficient, since the system infrastructure used to guarantee supply under peak hours is not completely used most of the time. According to the The US Department of Energy, 10\% of the whole generating capacity and 25\% of distribution capacity is used less than the 5\% of the time.

\begin{figure}[H]
  \centering
  \includegraphics[width=\columnwidth]{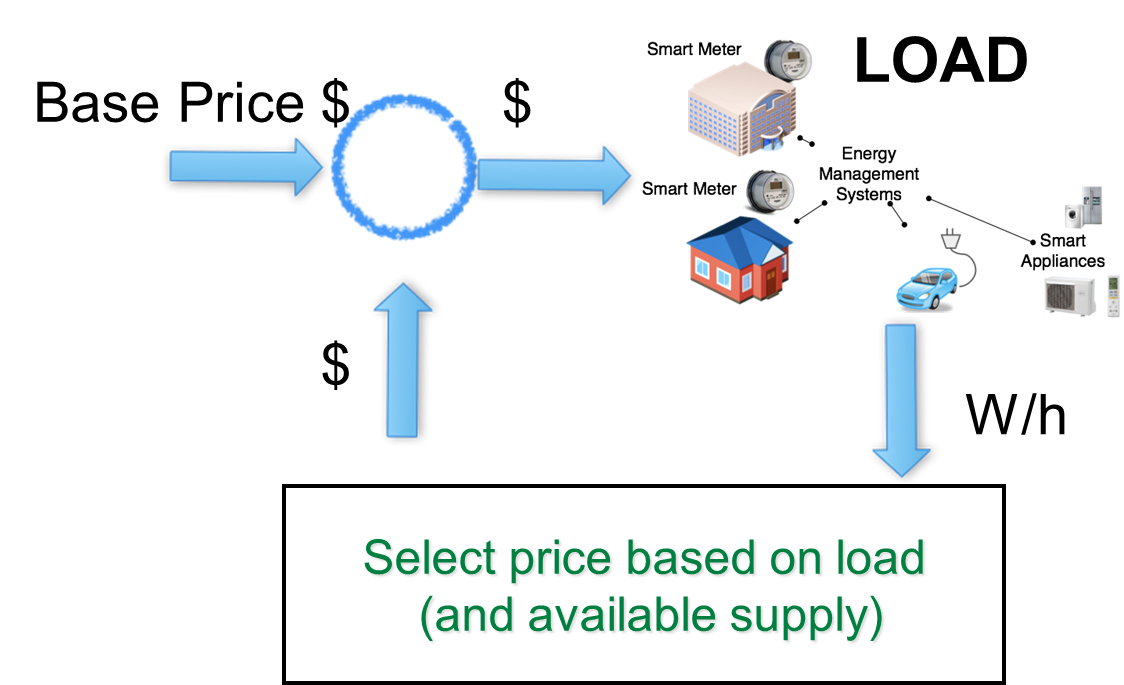}\\
  \caption{Real-time pricing algorithms try to control the load with price signals.}\label{DR-RTP}
\end{figure}

In an effort to increase the efficiency of the power grid, many retail-markets are expanding the use of demand-response programs. In their basic form, demand-response programs are a control problem where the control signal are the incentives (e.g., real-time pricing), or direct-load control (e.g., the utility directly controlling the set-points of air conditioning systems in specific cases) for consumers to reduce electricity consumption during peak hours and to shift this load to off-peak hours (Figure~\ref{DR-RTP}). Currently most of the electricity consumers leveraging demand-response programs are large commercial consumers, but the market is expanding more and more to smaller industries and even residential consumers. As the number of smart devices necessary to manage this market expands, the potential attack surface of the market also increases, and therefore we need to begin considering the potential impact of attackers that compromise devices and communication channels used in this market.

The security of demand response algorithms with real-time electricity pricing was recently explored by Tan et al.~\cite{tan2013impact}. In their work, they consider an attacker that has compromised a portion of the communication channels used to send price information to consumers, and then study the effects to the power system from \emph{scaling} and \emph{delay} attacks, where the prices advertised to smart meters are compromised by a scaling factor (so consumers use the wrong prices) and by corrupted timing information (so consumers  use old prices).  While this previous work is an important step for initiating the discussion on how to analyze the impact of attacks on real-time pricing, this research has limitations on the way it modeled the adversary by limiting attacks to scaling and delays. In addition this previous work did not discuss any security countermeasures against attacks. 

In this paper we extend the work of Tan et al.~\cite{tan2013impact} in several directions:
\begin{itemize}
\item Parametric adversary models (e.g., scaling or delay attacks) are a common assumption to keep a mathematical analysis of the problem tractable, but constraining the adversary this way is a deficiency in modeling realistic attackers which will not be subject to these  constraints. We model a more realistic attacker that can inject an arbitrary modification to the price received by the consumer, and is not constrained to scaling or delay attacks. 
\item Real-time pricing forms a closed-loop control system, and small modifications to the signals in the closed loop made by the adversary can be iteratively amplified by the feedback.  We use sensitivity analysis to identify the attack signals that will be amplified and the ones that will be attenuated by the control loop, and thus, we find the worst-possible attacks for any given bound on the maximum price deviation introduced by an attacker at any time instant. 
\item In addition to modeling and analyzing the impact of attacks that compromise the price signals, we also model the effect of attacks that compromise sensor signals (smart meter electricity consumption reports).
\item We propose countermeasures based on changing the parameter of the original controller by Tan et al. In addition, we propose an estimator and a new robust-control design that estimates the perturbation and computes a new price to attenuate the error between supply and demand caused by the attacker. We also introduce a low-pass filter as a solution to attenuate any high-frequency component of an attack, thus guaranteeing that our robust controller will minimize the difference between power generated and power consumed with, or without attack.
\item Finally, while the robust controller will minimize the impact of any attack, it will still be beneficial to notify the operator of the power grid of any potential indicator of an attack. Thus we propose an attack-detection algorithm and evaluate its effectiveness in a preliminary experiment to identify the properties of the detector for different controller parameters, and different attack frequencies.
\end{itemize}

\section{Related Work: Impact of Integrity Attacks in the Power Grid}

Our work falls within the scope of integrity attacks (or false-data injection attacks) to the sensor or control signals of a cyber-physical system. Integrity attacks have been proposed as a way to analyze the vulnerability of cyber-physical systems in general and the power grid in particular.  Injecting false data to state estimation algorithms used in bulk of the power grid was first proposed by Liu et al.~\cite{falsedata}, and similar integrity attacks were proposed for compromised smart meters trying to defraud the electric utility~\cite{mashima2012evaluating}.

The work on integrity attacks against bad data detectors for state estimation in the power grid has generated a significant body of follow up work; for example  D\'{a}n and Sandberg~\cite{Dan10}, consider a defender that can secure individual sensor measurements by, for example, replacing an existing meter with another meter with better security mechanisms such as tamper resistance or hardware security support. Kosut et al.~\cite{kosut} also extend the basic false data injection attack to consider attackers trying to maximize the error introduced in the estimate, and defenders with a new detection algorithm that attempts to detect false data injection attacks. Similar false-data injection attacks have been considered for specific devices in the power grid, such as integrity attacks against the Flexible Alternate Current Transmission System (FACTS)~\cite{FACTS,manimaran2011}, and Automatic Generator Control (AGC)~\cite{Esfahani2010ACC,Esfahani2010CDC,manimaran2010}. All this related work has targeted operational data of the power grid, and is not related to electricity markets.

Negrete-Pincetic et al.~\cite{gross09} were one of the first to study how false control signals can affect the social welfare of the electricity market. Related work by Xie et al.~\cite{xie} studied how false data injection attacks can be used to defraud bulk electricity markets by modifying Locational Marginal Prices (LMPs), and work by Jia et al.\cite{LiyanThomasTong12} studied how false meter data in the bulk of the power grid can be used to cause the largest errors in LMP estimation.
These integrity attacks have been studied in the bulk electricity market and specifically,  the estimation problem alone; most previous work does not consider how the control algorithm can be designed to minimize the impact of integrity attacks, or studied the feedback control loop behavior of the system under attack.

\section{Preliminaries}

\subsection{Demand Response Model}\label{sec:model}

We follow the real-time pricing model from Tan et al.~\cite{tan2013impact}. This model considers a market with consumers of electricity, suppliers of electricity, and a third party entity---an Independent System Operator (ISO)---with the goal of matching supply and demand by setting the market price for electricity.  The general assumption is that the ISO determines a clearing price $\lambda_k$ valid for the period of time $[k\cdot T, (k+1) \cdot T]$ (this is called an \emph{ex-ante} market) every $T$ hours (e.g., T=0.5h) and announces it to the suppliers and consumers.


The electricity demand is characterized by two components: a baseline electricity consumption $b_k$ that captures the electricity consumption that is independent of the pricing mechanism, and a price-responsive demand $w(\lambda_k)$, which captures the amount of electricity consumption that can be controlled by the pricing signal $\lambda_k$.  

The aggregated demand of all consumers is thus: 
$$
d_k(\lambda_k) = 
b_{k} + w(\lambda_k).
$$

For simulation purposes $b_k$ can be obtained from historical demand curves such as those from the New York ISO~\cite{NYISO}.

The Constant Elasticity of Own-price (CEO) has been commonly adopted to characterize the total price-responsive demand. The CEO model is defined by 
\begin{eqnarray}
w(\lambda_k)=D (\lambda_k)^\epsilon
\end{eqnarray}
where $D>0$ and $\epsilon\in(-1,0)$ are constants. The \emph{price elasticity of demand} is captured by $\epsilon$.

Similarly, for the supply of electricity, Tan et al., propose a linear regression between supply and cost, a model they validated from the Australian Energy Market Operator and the the electricity market in California. Under these assumption the supply of electricity can be modeled by the following equation:

\begin{eqnarray}
s(\lambda_k) = p \lambda_k + q,
\end{eqnarray}
where $p$ and $q$ are parameters estimated by the historical market data from the area of study.

\subsection{Control Objective}

%

The control objective of the ISO is to send price signals $\lambda_k$ to keep the error between supply and demand of electric power
$$
e_k = s(\lambda_k) - d(k,\lambda_k)
$$
close to zero for every time instant $k$. This can be seen as a control problem in which the system to be controlled is the outcome of a market, the control variable is the price signal $\lambda_k$ and the variable that can be measured is the error $e_k$.

The price signal $\lambda_k$ must be carefully designed because a direct feedback of the wholesale prices to the users might cause oscillations or even instability \cite{tan2013impact, mitter_2011}.

\subsection{Transfer Function Representation}

Transfer functions are a mathematical representation of linear difference (or differential) equations that allow us to represent the system in a compact way and to evaluate the performance of the system in therms of the frequency components of the control signals---recall that every time series has an equivalent representation (a one to one mapping) to a function in the frequency domain given by the Fourier transform. 

For our discrete-time system (where sensor and control actions are taken at given time steps $k$ separated by the sampling period $T$ (e.g., 30 minutes), the transfer function for the equations modeling the dynamics of the system can be obtained by using the z-transform (a transform similar to the Fourier transform).

In particular, we can define the transfer function of the price stabilization algorithm, the system, and the observation mechanism as $G_c(z)$, $G_p(z)$, and $H(z)$, respectively.

To express these transfer functions it is necessary to approximate the dynamics system at the operation point $\lambda_0$ to a linear system. Hence, following Tan~et~al.~\cite{tan2013impact} we make the following approximations with the Taylor polynomials of the supply $s()$ and demand $w()$:
\begin{eqnarray*}
s(\lambda) & \simeq  \dot{s}(\lambda_0) (\lambda -\lambda_0) + s(\lambda_0) \\ &  =  \dot{s}(\lambda_0) \lambda + s_0 \\
w(\lambda) & \simeq  \dot{w}(\lambda_0) (\lambda -\lambda_0) + w(\lambda_0) \\ &  =  \dot{w}(\lambda_0) \lambda + w_0
\end{eqnarray*}
where $\dot{f}=\frac{df}{d\lambda}$, and where we define the constant (or endogenous) terms with $s_0 = s(\lambda_0) - \lambda_0 \dot{s}(\lambda_0)$ and $w_0 = w(\lambda_0)- \lambda_0 \dot{w}(\lambda_0)$.

Therefore, the transfer functions can be defined as
$$
G_s(z) = \dot{s}(\lambda_0) = p,
$$
with initial  condition $s_0$ and
$$
G_w(z) = \dot{w}(\lambda_0) = D \epsilon (\lambda_0) ^{\epsilon-1} ,
$$
with initial  condition $w_0$.

The outcome of the market can be expressed as $G_p(z) = G_s(z) - G_p(z)$.

\subsection{Control Algorithm for Setting Prices}

The price setting control algorithm depends on the previous price $\lambda_{k-1}$ and the observed error at the current time step $e_k$. If $e_k$ is negative, it means that there was more power demanded than supplied, and thus the price will increase (to motivate consumers to decrease consumption), while if $e_k$ is positive, then the price will decrease. The precise amount of increase and decrease of the prices at each time step should be selected carefully as inadequate price updates can make the system unstable. Tan et al. found that when we design a proportional gain $\eta \in (0,1)$ in the following price-setting algorithm:
$$
\lambda_k = \lambda_{k-1} - \frac{2 \eta}{ \dot{s} (\lambda_0) - \dot{w}(\lambda_0) } e_k,
$$
The system will remain stable. $\eta$ is in fact an important design parameter for the control algorithm, and as we will show, it can also determine the impact to the resiliency of the system under attacks. When properly selected, it can also attenuate the impact of attacks.

Assuming an observation device characterized by a one-step delay transfer function: $H(z) =z^{-1}$, this price control mechanism can be represented by a transfer function as
$$
G_c(z) = \frac{2\eta}{ \dot{s} (\lambda_0) - \dot{w}(\lambda_0) } \frac{1}{1-z^{-1}}.
$$

\section{Attacker Model}

In contrast to one-shot attacks, where the attacker provides false information only once~\cite{teixeira2014_attack,falsedata}, in this work we consider that an attacker compromises a device or a communication channel,  and has the capability to add false information at any moment and---more importantly---repeatedly over a long period of time.  

For example, most of the work on false data injection in state estimation finds a value $d$ to insert at an arbitrary point in time~\cite{falsedata}, however, this  previous work does not consider the evolution of the system dynamics over time.  In this context, the question we would like to pose from an adversarial point of view is the following: 

\begin{itemize}
\item What is the worst attack time series $d_k$ that can affect the system while keeping some bounds (prices will be bound by some maximum and minimum values: $\forall k~~~ d_k \in [d_{min}, d_{max}]$.
\end{itemize}

Tan et al.~\cite{tan2013impact} proposed an adversary model where one attacker compromised the pricing communication channel between the ISO and a percentage $\rho$ of consumers. They considered delay attacks and scaling attacks. 

In a delay attack, the compromised price is an old version of the price, i.e., $\hat{\lambda}_k=\lambda_{k-\tau}$, and in a scaling attack, the compromised price is a scaled version of the true price, i.e., $\hat{\lambda}_k=\gamma \lambda_k$.

While the attacks defined above can be easily analyzed from a theoretical point of view, it is not clear why an attacker who has compromised a communications channel will select to launch these attacks when she has the flexibility of sending any arbitrary time series $\hat{\lambda}_k$ she wants, even one that bears no resemblance to the original time series $\lambda_k$.  

Furthermore, scaling attacks and delay attacks are not strategic, and do not seek to maximize any objective function from the adversary.  In this work we follow the generic and more powerful adversary model introduced by the false data injection paper~\cite{falsedata}, and we expand it to consider a time series. In particular, we model a compromised communication channel as $\hat{\lambda}_k=\lambda_k+d^a_K$, where $\hat{lambda}_k$ is the price information received by the victim, and $d^a_k$ is an arbitrary time signal that can take any value.  It is clear now that scaling attacks and delay attacks are simple subsets of this new attack because for every scaling or delay attack possible producing a false price information $\hat{\lambda}_k$, there exists an arbitrary time signal $d_k^a$ that will produce the same price $\hat{\lambda}_k$ received by the victim.

The question we now face is how to determine a strategic attack time series $d_k^a$ that will try to cause as much damage as possible (i.e., that will try to maximize the mismatch between power generated and power consumed).  One of our key insights into tackling this problem is the fact that for every time series, there is a one-to-one correspondence of the time series and its frequency (Fourier transform) representation.  Therefore, instead of attempting to analyze the worst time series $d_k^a$ in time, we identify the worst-possible attacks in frequency space. 

In order to provide a mathematical tool that enables us to quantify the impact of the attack, we use  sensitivity analysis. 
Sensitivity functions  have been widely used to analyze the impact of  external disturbances or parameter changes on the output  of a feedback system. In systems and control theory, it is well known that feedback can attenuate or amplify disturbances; therefore, using the frequency representation of the system (the transfer function),  it is possible to obtain the sensitivity function  and observe the response of the system to a perturbation of a specific frequency  $\omega$.~\cite{doyle2013feedback}.



In this work we focus our attention on two types of additive attacks: i) additive attack in the price information, and ii) additive attack in the sensor information. Each type of attack produces different consequences to the system.

In the next section we give the formal incorporation of the attacks against pricing signal, and in the section after that we use sensitivity analysis to identify the impact of the attacks.


\begin{figure*}
  \centering
  \includegraphics[width=0.9\textwidth]{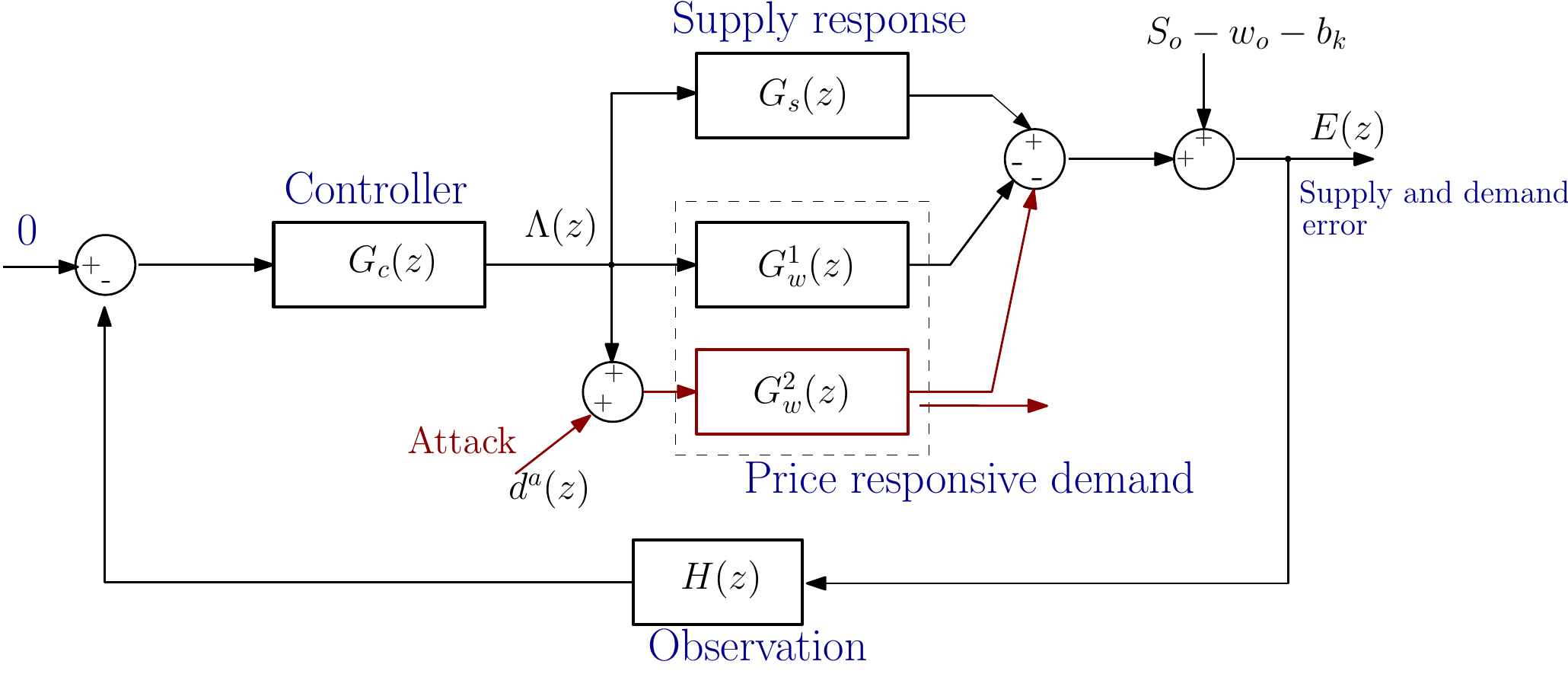}\\
  \caption{Block diagram of the real-time pricing model under attack.}\label{disturbance_model}
\end{figure*}

\subsection{Incorporating the Attack into the Real-Time Pricing Model}

We assume  that an amount $\rho$ of communication channels are compromised, and each of these consumers receives the price value $\hat \lambda_k=\lambda_k+d^a_k$, where
$d^a\in \mathbb{R}$ corresponds to the additional false information. 


It is necessary to identify how the inclusion of this attack affects the system representation of the real-time problem. In particular, we need to identify how the attack changes the transfer functions of the model (i.e., we need to characterize the new transfer functions $G_w^1(z)$ for the consumers who are unaffected, and $G_w^2(z)$ for the consumers who receive false information, as shown in Figure \ref{disturbance_model}.)

Let us consider the price response demand based on the CEO model for the set of compromised nodes $\rho w_k(\lambda_k,d^a_k)=\rho D(\lambda_k+d^a_k)^\epsilon$. In order to linearize this model it is necessary to assume that $|d_k|<<\lambda_k$ and $\lambda_k>0$. As we will discuss towards the end of the paper (the attack-detection formulation), this is a perfect assumption for an attacker that wants to minimize its chances of being detected (by causing small changes to the price $|d_k|<<\lambda_k$) but at the same time wants to find the best way to find a small signal deviation that will maximize the potential damage to the system. 

The linearized model is described by:
\begin{eqnarray*} 
w(\hat \lambda_k) &= & \rho w(\lambda_o+d^a_o)+ \\ 
 & & \rho \dot w(\lambda_o+d^a_o)\left(\lambda_k+d^a_k-\lambda_o-d^a_o\right)+ \\ & & (1-\rho)(w(\lambda_o)+\dot w(\lambda_o)(\lambda_k-\lambda_o))
\end{eqnarray*}

We can group the price-independent terms with $b_k$ (the baseline consumption of electricity that is independent of the price),  and then also group  the price-dependent components for the transfer functions.  
\begin{eqnarray}
G_w^1(z)=(1-\rho)\dot w(\lambda_o),
\end{eqnarray} then  corresponds to the transfer function of consumers who receive unmodified price information, and 
\begin{eqnarray} 
G_w^2(z)=\rho\dot w(\lambda_o+d_o),
\end{eqnarray} corresponds to the transfer function of the victims. Under the assumption that $|d_k|<<\lambda_k$, we can neglect the term $d_o$ in the linearization, such that 
\begin{eqnarray}
G_w^2(z)=\rho\dot w(\lambda_o).
\end{eqnarray}

\section{Sensitivity Analysis}

\begin{figure*}
  \centering
  \includegraphics[width=0.7\textwidth]{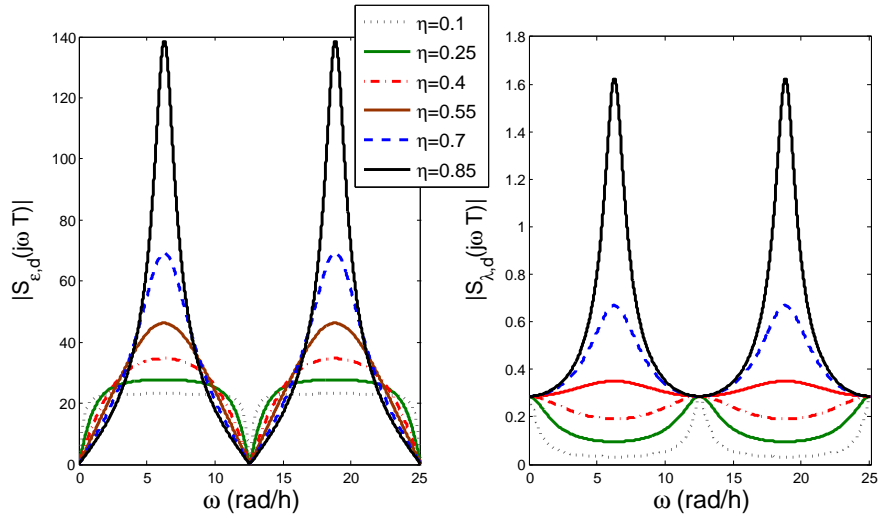}\\
  \caption{Left: Sensitivity of the error $E(z)$. Right: Sensitivity of the Price $\lambda(z)$. We can see that while the attack always amplifies the error between power generated and consumed, the price signals are actually attenuated (except for $\eta=0.85$). This sensitivity analysis uses parameters:  $\rho=0.5, p=31,q=917$, and $T=0.5$h. The baseline consumption is $b=400$ MW, which is proportional to 1 million households, and the base demand of each consumer is $b_i\in[2.8,4.6]KW$.}\label{Fig_sens_function1}
\end{figure*}

The sensitivity function models how one input to the system (in our case the attack) affects another signal in the system (we are mostly interested to see how the attack affects the error in power generated minus the demand, and to also see the impact on the prices).

We start by looking at the impact that a disturbance $d^a(z)$ (in the frequency space) can have on the error $\mathcal{E}(z)$. In particular, the sensitivity function  for these two time series (denoted as $S_{\mathcal{E},d}$) is the ratio $\mathcal{E}(z)/d^a(z)$:
\begin{eqnarray}\label{eq_sed}
S_{\mathcal{E},d}&=&-\frac{G_w^2(z)}{1+G_c(z)H(z)G_p(z)} \nonumber \\ &=&-\frac{\rho \dot w(\lambda_0)(z-1)}{(z-1+2\eta)}.
\end{eqnarray}

As stated before, our interest is to analyze the effects of an additive attack in the frequency domain. We denote the angular frequency as $\omega$. 
We then replace $z=e^{j\omega T}$ for $T$ being the sampling period (the time interval between updating the sensor measurements and the prices). It is important to notice that the maximum frequency that an attacker can generate is limited by the sampling period, such that $\omega_{max}=\pi/T$. For instance, if the sampling period is $T=0.5$~hours, then $\omega_{max}=2\pi$. 

In order to observe the effects an attack time-series with different frequency components in the output error $\mathcal{E}(z)$, we obtain the expression $|S_{\mathcal{E},d}(e^{j\omega})|$ for $\omega$, the disturbance frequency:
\begin{multline*}
 |S_{\mathcal{E},d}(e^{j\omega})|=  \\
  \frac{|\rho \dot w(\lambda_0)|\left(\sin^4(\omega/2)-2\eta\sin^4(\omega/2)+\eta^2\sin^2(\omega/2)\right)^{1/2}}{\left(\sin^2(\omega/2)-2\eta\sin^2(\omega/2)+\eta^2\right)}
\end{multline*}

\begin{figure}[h]
  \centering
  \includegraphics[width=\columnwidth]{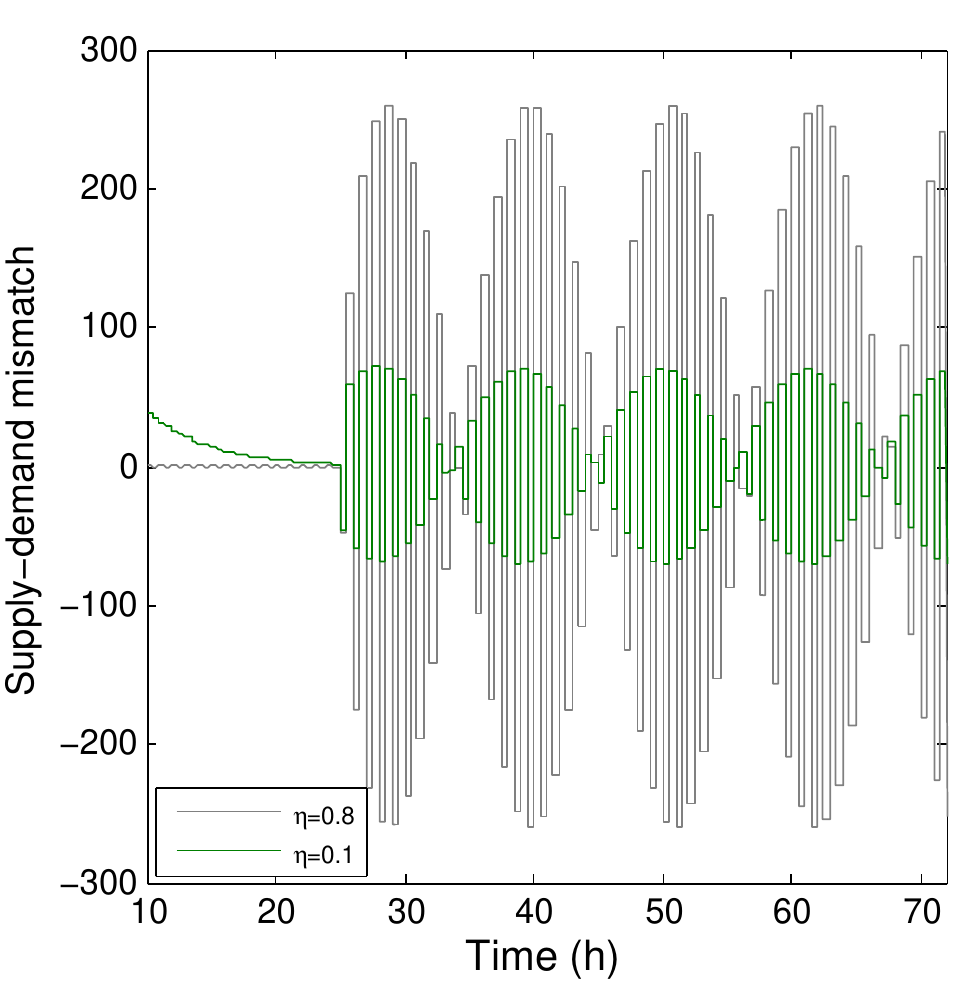}\\
  \caption{A smaller control parameter $\eta$  will be able to attenuate the impact of high-frequency attacks; however this will come at the cost of longer convergence times.}\label{convergence}
\end{figure}

From this equation we can see that the percentage of compromised channels $\rho$ has a scaling effect on the sensitivity of the system. Moreover, the selection of the control parameter $\eta$ proposed by Tan~et~al. is fundamental for attenuating the effects of the attack. The left side of Figure \ref{Fig_sens_function1} shows how the attack can be amplified (or attenuated) as a function of the frequency of the attack signal. Clearly, the impact the supply-demand mismatch $\mathcal{E}$ is severe for most frequencies; however, we can also see how the control parameter $\eta$ can be selected to attenuate the impact of high-frequency signals: smaller values of $\eta$ will minimize the impact of high-frequency components of the attack time-series---this comes at the cost of a slower control action (as seen in Figure~\ref{convergence}) which might not be a bad idea, as changes in prices will remain small, giving consumers more predictability in their electricity consumption habits. 

Recall that if the output $\mathcal{E}$ is different from zero, then  there is over demand or over production of electricity, which can affect considerably the system (resulting in large frequency changes). Even if the price variations are small, the output amplifies the disturbance. There is a trade off between the $\eta$,  $\rho$, and the frequency of the disturbance. An attacker can easily take advantage of this fact, and introduce intelligently false data to a portion of the users. This information can be of small amplitude, and hardly detected; however, the effects on the output can be catastrophic.

We can also obtain the sensitivity function with respect to the price. This function reveals how the attack modifies the real price calculated by the ISO.  The function is  described by
\begin{align}
S_{\lambda,d}(z)&=-\frac{G_c(z)G_w^2(z)H(z)}{1+G_c(z)H(z)G_p(z)} \nonumber \\
&=-\frac{2\eta \rho\dot w(\lambda_0)}{(\dot s(\lambda_0)-\dot w(\lambda_0))(z-1+2\eta)},
\end{align}
and looking at the magnitude of the frequency components we obtain:
\begin{multline*}
|S_{\lambda,d}(e^{j\omega})|=  \\  \frac{|\eta \rho\dot w(\lambda_0)|}{(\dot s(\lambda_0)-\dot w(\lambda_0))\left(\sin^2(\omega/2)-2\eta \sin^2(\omega/2)+n^2\right)^{1/2}}.
\end{multline*}
The left side of Figure~\ref{Fig_sens_function1} shows the sensitivity function with respect to the price for different values of $\eta$, and $\rho=0.5$. With this selection of $\rho$, the real price changes produced by the attack are attenuated for all $\eta$.

\subsection{Applying Lessons Learned in the Frequency Domain to the Time Domain}
\begin{figure}[H]
  \centering
  \includegraphics[width=\columnwidth]{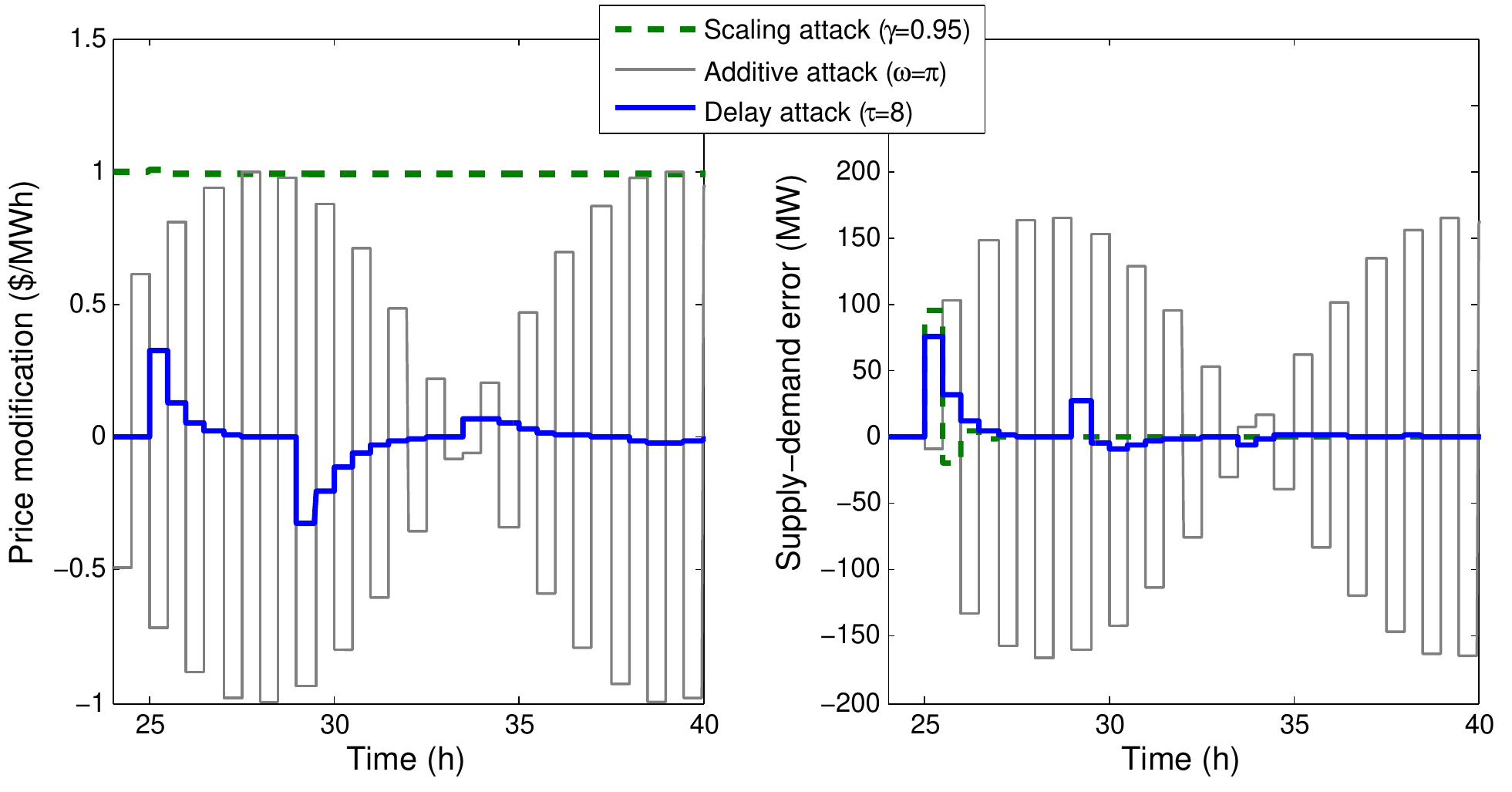}\\
  \caption{Modification in the price for the scaling attack with a scale parameter $\gamma=0.95$, delay attack with a delay $\tau=8$, and the additive attack $d_k^a=\sin(2\pi kT)$ }\label{fig_example}
\end{figure}

\begin{figure*}
  \centering
  \includegraphics[width=0.7\textwidth]{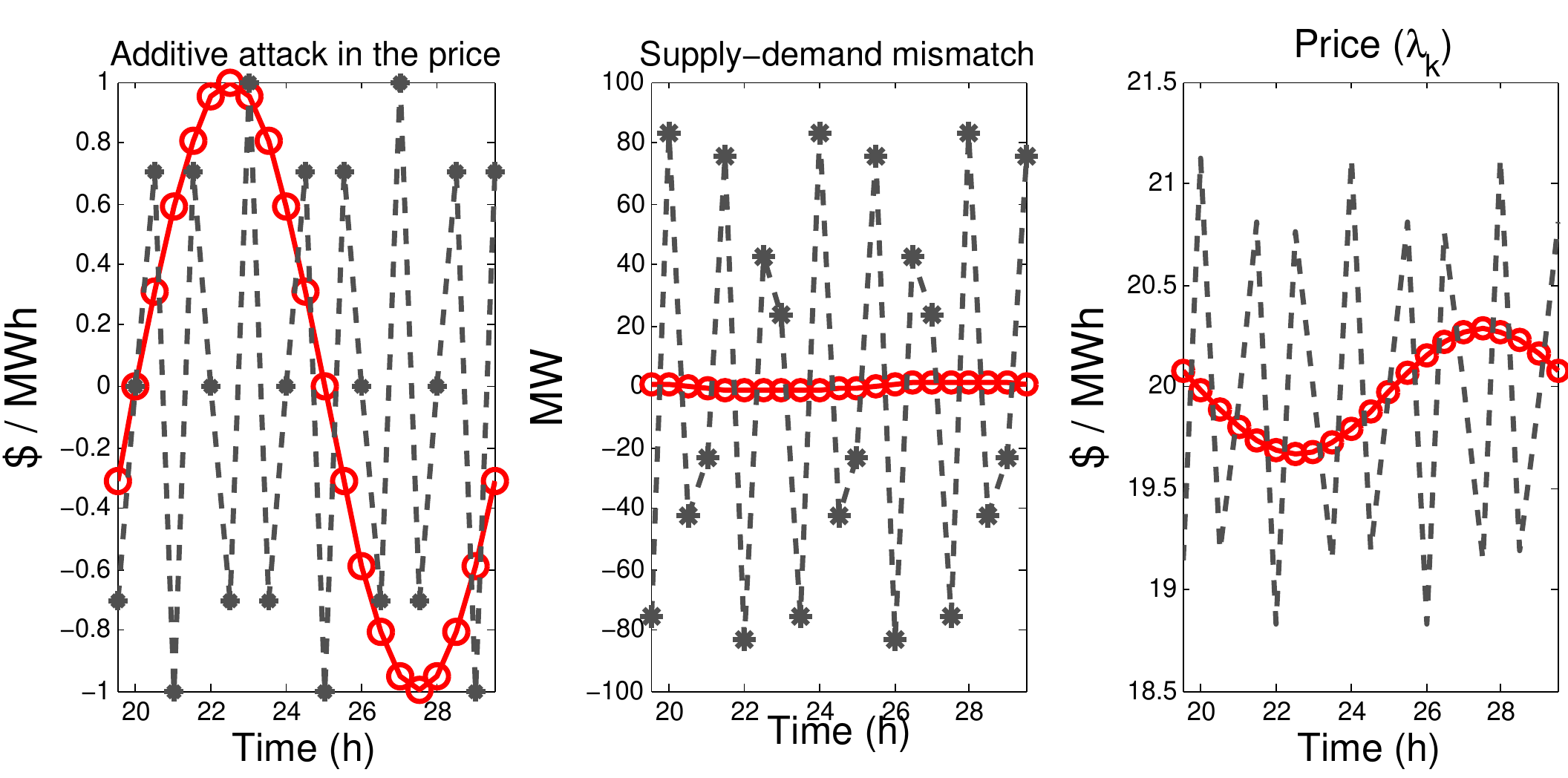}\\
  \caption{Effects in the supply-demand mismatch (middle) and the price (right) for two attacks i) $d_k^a=\sin(3/2\pi T k)$ and ii)$d_k^a=\sin(\pi T k/5)$ for $\eta=0.8$}\label{fig_example1}
\end{figure*}

\begin{figure*}
  \centering
  \includegraphics[width=0.7\textwidth]{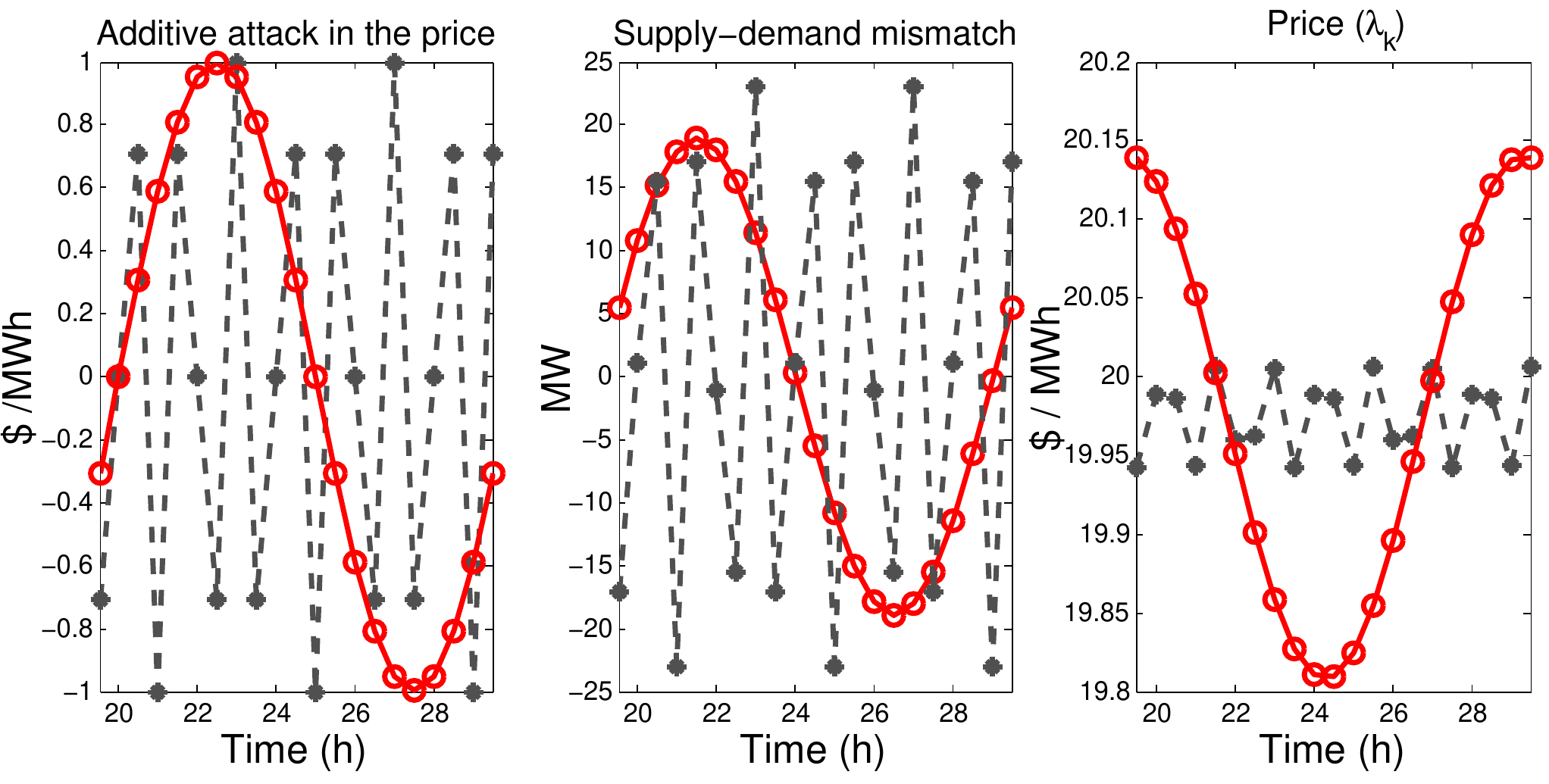}\\
  \caption{Effects in the supply-demand mismatch (middle) and the price (right) for two attacks i) $d_k^a=\sin(3/2\pi T k)$ and ii)$d_k^a=\sin(\pi T k/5)$ for $\eta=0.1$}\label{fig_example2}
\end{figure*}

Now that we have gained some insight into how the ``frequency components'' of a time series can affect the system, we look again at the ``time domain'' to apply these lessons in the analysis of attacks.

First we take a look at how the attacks proposed in previous work (scaling and delay attacks) compare to attacks with a frequency designed to maximize the error between generated and consumed power. Figure~\ref{fig_example} shows a typical example of a the effects of a scaling attack, a delay attack, and the attack targeting the frequency where the maximization of the error is maximum. The left hand side of the figure shows three different attack time series: the green signal is the scaling attack, the blue signal is the delay attack, and the black signal is the new attack designed with our sensitivity function analysis. The right hand side of Figure~\ref{fig_example} shows how previously proposed attacks generate a much smaller error than the attack designed with the help of the sensitivity function.

We now look at attacks of different frequencies and their effect on both: (1) the error in generated and consumed power, and (2) the price signal. 

{Figure~\ref{fig_example1}} shows a high-frequency attack (black) and a low-frequency attack (red) on the left. The control algorithm is using $\eta=0.8$ and therefore we can see a large error magnification caused by this control parameter (as predicted by Figure~\ref{Fig_sens_function1}). Similarly, the price signal is also amplified for the high frequency attack (as can be seen by the figure on the right). 

Figure~\ref{fig_example2} shows a high-frequency attack (black) and a low-frequency attack (red) on the left. The control algorithm is using in this case the parameter $\eta=0.1$, and it can be seen (in the middle figure) how the impact of the error between supply and demand is attenuated when compared to Figure~\ref{fig_example1}. The other interesting thing to observe on the figure at the right is that (as predicted by Figure~\ref{Fig_sens_function1}) the price signal is attenuated for high frequencies when we use small $\eta$.

\section{Modeling Attacks on Sensors}

Previous work has only considered integrity attacks to the price signals, but the sensors (or smart meters) can also be compromised and can be used to send false electricity consumption reports to the controller.   This new attack model requires a new mathematical analysis of the attacks. 

\begin{figure}[H]
  \centering
  \includegraphics[width=\columnwidth]{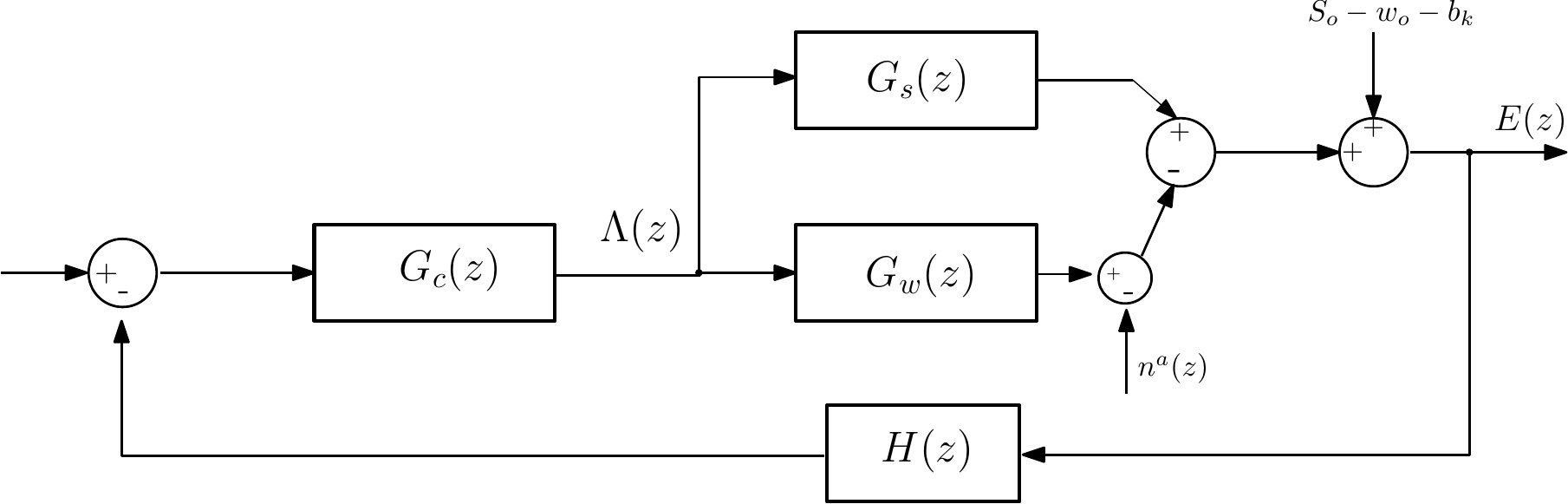}\\
  \caption{Block diagram of the real-time pricing model with an attack $n^a$ on the sensor values.}\label{fig_sensor_disturbance_scheme}
\end{figure}

Now we assume that the  attack occurs in the information that each consumer sends to the ISO, where $N$  sensors are compromised (Figure \ref{fig_sensor_disturbance_scheme}). We can observe that the main difference between attacking price signals (i.e., control commands) and sensor signals, is the fact that sensor signals are going to be aggregated in this case, and therefore we do not need to model two different transfer functions for compromised consumers, and uncompromised consumers (as we had to do when the price signal was attacked).

We define $n^a_k$ as the attack over one sensor, and study the sensitivity for one attack, and due to the linearity of the model and the assumption of homogeneous attacks, we scale the analysis by a factor $N\in\mathbb{Z}_+$, which is the number of compromised sensors.

The sensitivity function that relates the output $\mathcal{E}(z)$ with respect to the sensor additive attack $n^a$ is given by
\begin{eqnarray}
S_{\mathcal{E},n}&=&  -\frac{N}{1+G_p(z)H(z)G_c(z)} \nonumber \\ &=&-N\frac{z-1}{z-1+2\eta}.
\end{eqnarray}
Evaluating $z=e^{j\omega}$, we obtain the frequency response of the sensitivity function as
\begin{multline*}
|S_{\mathcal{E},d}(e^{j\omega})|= \\ \frac{N\left(\sin^4(\omega/2)-2\eta\sin^4(\omega/2)+\eta^2\sin^2(\omega/2)\right)^{1/2}}{\left(\sin^2(\omega/2)-2\eta\sin^2(\omega/2)+\eta^2\right)}
\end{multline*}

\begin{figure}[htbp]
\centering
\includegraphics[width=\columnwidth]{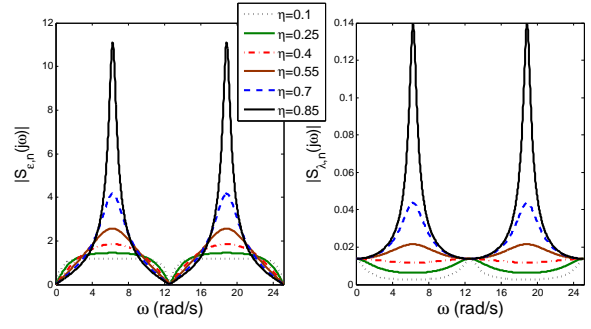}
\caption{ Sensitivity with respect to the supply-demand error (left) and price (right) for only one compromised sensor.}
\label{Fig_sensor_mod}
\end{figure}

Similarly, we evaluate the effects of the price variations provoked by the false sensor information:
\begin{align}
S_{\lambda,n}(z)&=-N\frac{Gc(z)H(z)}{1+G_p(z)H(z)G_c(z)} \nonumber \\ &=N\frac{2\eta}{(\dot s(\lambda_0)-\dot w(\lambda_0))(z-1+2\eta)}
\end{align}
\begin{multline}
|S_{\lambda,n}(e^{j\omega})|= \\ \frac{N|\eta|}{(\dot s(\lambda_0)-\dot w(\lambda_0))\left(\sin^2(\omega/2)-2\eta \sin^2(\omega/2)+n^2\right)^{1/2}}.
\end{multline}

Figure~\ref{Fig_sensor_mod} shows the graphical representation of both sensitivity functions.  Clearly, the sensitivity functions with respect to the sensor attack are scaled versions of the sensitivity with additive attack in the price. Therefore, if the additive attack in the price occurs with $\rho=1$, for a total number of consumers $N_T>>|\dot(\lambda_o)|$,  the effects of the same attack in all the sensors ($N=N_T$) will produce a larger impact.

\begin{figure}[htbp]
\centering
\includegraphics[width=\columnwidth]{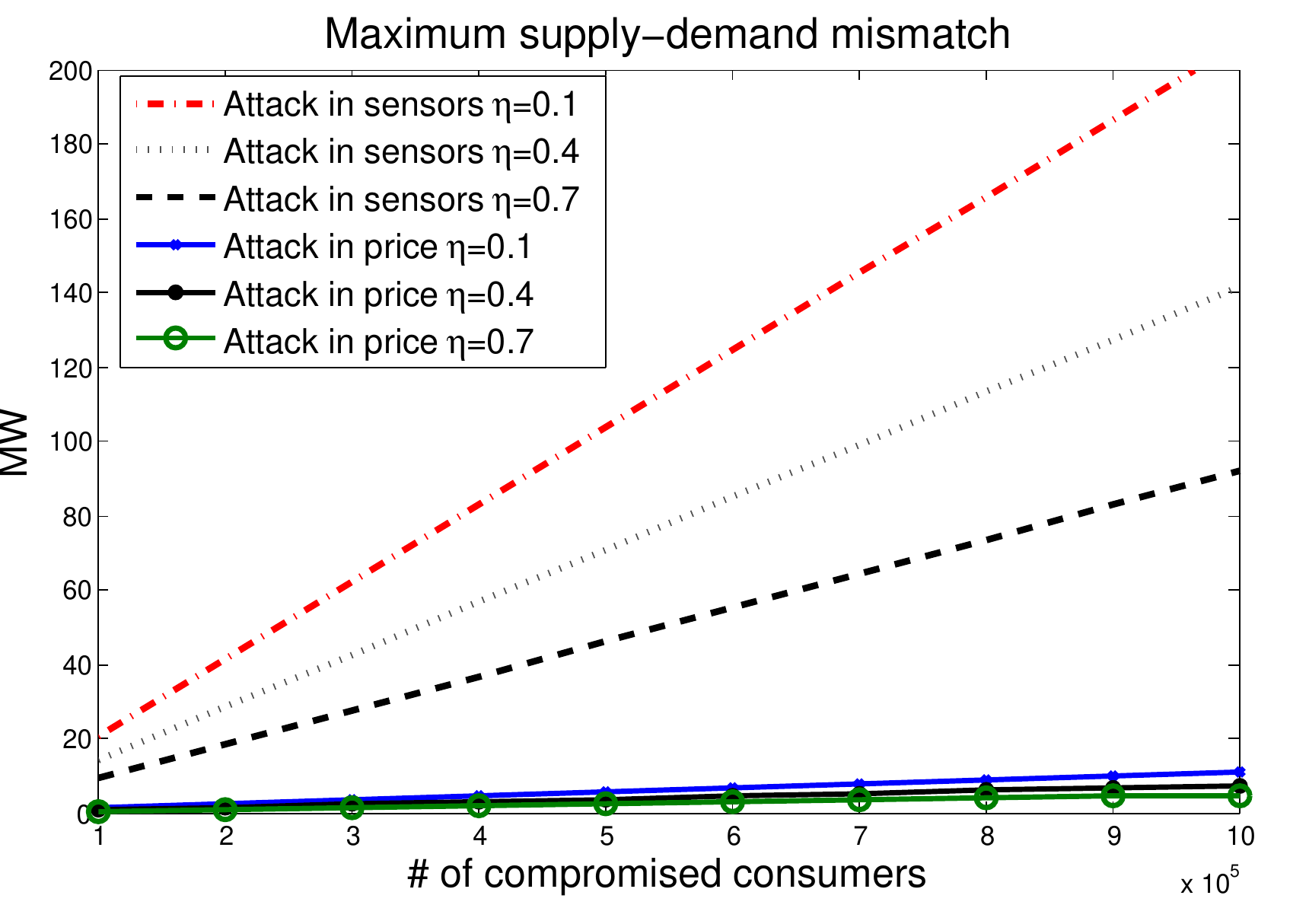}
\caption{Comparison between the additive attack over the price information and over the sensors. }
\label{Fig_compare}
\end{figure}

In order to illustrate the different impacts for both types of attacks (control signals vs. sensor signals) we assume a total number of consumers $N_T=1000000$. We analyze two different cases: i) an  attack in the price information with $d_k^a=0.25\sin(\pi k/2)\$/MW$, ii)  an  attack on the sensor measurements with $n^a_k=0.2\sin(\pi k/2)kW/h$.

Figure~\ref{Fig_compare} shows the maximum value of the output $\mathcal{E}$ when a disturbance of the form $d_k=0.25\sin(\pi k/2)$ is introduced to the price value and to the sensors, for different values of $\eta$, and for different amount of compromised consumers (for both types of attacks). We can see that for the same number of communication channels compromised, the attacks can be actually much worse if the attacker decides to compromised sensors vs.~compromising the control signals.

%
%
%
%
%
%

\section{Designing an Attack-Resilient Controller}


Previous work only studied the effects of the attack, but did not propose new control mechanisms to mitigate possible attacks. We know discuss how we can start designing attack-resilient controllers. 

In order to design an attack-resilient controller, we can leverage the fact that the ISO has historical data showing the behavior of the system which can be used for learning the dynamics (parameters) of the system. Whenever the controller commands do not have the expected effect, or when the sensor signals to not reflect the normal evolution of the system we can try to identify these problems and design a controller that minimizes the impact of price or sensor attacks. 

As the attack are unknown inputs into the system, we can use a type of disturbance estimators. Disturbance observers have been studied in literature but we focus our attention in the one introduced by Kim et al.~\cite{kim2013968} for discrete-time systems.

We assume that the ISO possesses the information about the supply-demand error $\mathcal{E}_{k-1}$ and we try to detect an attack using the observer (an observer is another name for a ``state estimator''). 

We first present the attack-resilient controller for a general discrete-time system, and in the next section we show how to apply it to our real-time pricing model.

Let us consider a generic linear discrete-time system  for a sampling period $T>0$ of the form
\begin{align}\label{eq_discrete_sys}
x_{k+1}&=Ax_k+Bu_k+\Gamma d_k\nonumber\\
  y_k&=Cx_k
\end{align}
where $x_k\in\mathbb{R}^n$, $u_k\in\mathbb{R}^m$, $d_k\in\mathbb{R}^q$, and $y_k\in\mathbb{R}^l$ are the state variable, the control input, the disturbance, and the measurement output, respectively. The matrices $A,B,\Gamma,C$ are of adequate dimension.\\

For $d_k=(d_k^1,\ldots d_k^q)$, the disturbance is slowly time-varying, such that
$d_{k+1}^i-d_k^i<T\mu_i$, $\forall i=1,\ldots,q$.
Given a $K\in{\mathbb{R}^{q\times n}}$ and $C=I_n$, the observer is described as follows
\begin{align}
z_{k+1}&=z_k+K\left((A-I_n)x_k+Bu_k+\Gamma\hat d_k\right)\nonumber\\
\hat d_k&=Kx_k-z_k
\end{align}
Under the assumption that $\Gamma$ is invertible, we can  choose $K=(I_q-\Phi)\Gamma^{-1}$ for $\Phi=[\phi_1,\ldots,\phi_q]^\top$, and $\phi_i\in(-1,1)$.  The estimation error  is then bounded by
\[e_{\infty}=\frac{T\mu_i}{1-|\phi_i|}
\]
for $\phi\in(0,1)$, and $\mu_i>0$.

However, as the ISO possesses only past information about the state  (i.e., $x_{k-1}=\mathcal{E}_{k-1}$) the estimator has to be slightly modified in order to estimate the disturbance using only $x_{k-1}$.  As a consequence, the estimation is always a delayed version of the real disturbance. Therefore, the modified estimator is given by
\begin{align}\label{eq_mod_estimator}
\hat d_{k}&=Kx_{k-1}-z_{k-1} \nonumber\\
z_{k+1}&=z_{k-1}+K\left((A-I_n)x_{k-1}+Bu_k+\Gamma\hat d_k\right)
\end{align}

\subsection{Estimation of Price Attacks}
Let $G_p=\dot s(\lambda_0)-\dot w(\lambda_0)$ and $d_k=d_k^a$ to simplify notation. We can write the feedback real-time pricing problem using a   discrete-time state  space representation as follows
\begin{align}\label{eq_RTP_state}
\mathcal{E}_{k+1}&=G_pu_k-\rho\dot w(\lambda_0)d_k\nonumber\\
y_k&=\mathcal{E}_k
\end{align}
Note that comparing Equation~(\ref{eq_RTP_state}) with Equation~(\ref{eq_discrete_sys}), we have $A=0$, $B=G_p$, $\Gamma=-\rho\dot w(\lambda_0)$, $x_k=\mathcal{E}_k$ and $u_k=\lambda_k$.

Note that to compute the state estimation, it is necessary to know $\Gamma$, which means that we would need prior knowledge about the amount of compromised nodes. Obviously, this requirement seems unrealistic as $\rho$ will remain unknown to the defender; however, we can exploit a very interesting property  of the estimator we found to perform state estimation without knowing $\rho$, as stated in the  following proposition.

\textbf{
\begin{prop}\label{prop1}
Let us consider the disturbance estimator described in (\ref{eq_mod_estimator}) for the real-time pricing model in Equation (\ref{eq_RTP_state}). The rate of change of the disturbance $\Delta_{k}=d_k-d_{k-1}$ is bounded such that  $\Delta d_{k}\leq T\mu$ for some constant $\mu$ and $T$ the sampling period.  We define  $\hat \Gamma$ as an approximate value of $\Gamma$ and  $\hat e_k=\Gamma d_k-\hat \Gamma\hat d_k$ as an error between the real effect of the disturbance and its estimate. If  $K=\hat \Gamma^{-1}(1-\phi)$ for $\phi\in(-1,1)$, the error converges and is bounded by
\[ |\hat e_{\infty}|\leq \frac{2|\Gamma|  T\mu}{1-|\phi|} \] 
\end{prop}
}
\emph{Proof:}\\
The error evolution is 
\footnotesize{
\begin{align*}
\hat e_{k+1}&=\Gamma d_{k+1}-\hat \Gamma \hat d_{k+1}\\
            &=\Gamma d_{k+1}-\hat \Gamma(Kx_k-z_k)\\
            &=\Gamma d_{k+1}-\hat \Gamma K(G_pu_{k-1}+\Gamma d_{k-1})\\
            &\text{  }+\hat \Gamma (z_{k-2}-Kx_{k-2}+K\Gamma pu_{k-1}+K\hat \Gamma \hat d_{k-1})\\
            &=\Gamma d_{k+1}-\hat \Gamma K\Gamma d_{k-1} -\hat \Gamma d_{k-1}+\hat \Gamma  K \hat \Gamma \hat d_{k-1}\\
            &=\Gamma d_{k+1}-\Gamma d_{k-1}-\hat \Gamma  K(\Gamma d_{k-1}-\hat \Gamma \hat d_{k-1}) -\hat \Gamma d_{k-1}+\Gamma d_{k-1}\\
            &=2\Gamma \Delta d_{k+1}+(1-\hat \Gamma K)\hat e_{k-1}
\end{align*}
}
\normalsize
As $K=(1-\phi)/\hat \Gamma $, in the equilibrium when $\hat e_{k+1}=\hat e_{k-1}$, $\hat e_{\infty}$ is bounded by
\[|\hat e_{\infty}|=\frac{2\Gamma \Delta d_{k+1}}{1-|\phi|}\leq \frac{2\Gamma T\mu}{1-|\phi|}
\]
\hfill $\blacksquare$
\begin{remark}
If the portion of compromised nodes is identified, then the estimation error $e_k=d_k-\hat d_k$ converges and is bounded by \[e_{\infty}\leq \frac{2 T\mu}{1-|\phi|}\]
\end{remark}

\subsection{Estimation of Sensor Attacks}

Similar to the previous case, estimating the disturbance $n^a$ does not require prior knowledge of $\rho$ due to the fact that the attack modifies the information that consumers provide about its consumption and this affects directly the demand. 

The state estimation can then be given by
\begin{align}
z_{k+1}&=z_{k-1}+K\left(-\mathcal{E}_{k-1}+G_p\lambda_k+\Gamma\hat n_k^a\right)\nonumber\\
\hat n_k^a&=K\mathcal{E}_{k-1}-z_{k-1}
\end{align}
where $\Gamma=-1$.
As it was proven before, $\hat e_k$ is bounded independent of $\hat \Gamma$. This fact will be useful  to detect attacks without knowing its exact location, i.e., without knowing if the attack is modifying the price or the sensor information, and we can do this using the same estimator (of course if the attacker controls both: all price signals, and  all sensor signals then there is nothing we can do as we have lost any hope of getting situational awareness from the system). 

Before introducing the proposed detection mechanism, we will show how to improve  disturbance rejection of the system using the estimator.

\subsection{Robust Control Algorithm}

It is possible to modify the disturbance rejection using an add-on compensator in the controller of the form
\[u_k=u_{nom}-B^{-1}\Gamma\hat d_k=\lambda_k-Gp^{-1}\hat\Gamma\hat d_k\]
where $u_{nom}$ is the controller under normal conditions.

The mismatch between the supply and the demand is described by
\[\mathcal{E}_{k+1}=G_p\lambda_k+\Gamma d_k-\hat \Gamma \hat d_k\]. 

Clearly, if $\hat e_k$ is small, disturbances are attenuated.

Including the robust controller in the system produces an improvement in the estimation,  leading to the following result.
\textbf{
\begin{prop}\label{prop2}
For the RTP system under  additive attack, and the proposed robust controller $\hat \lambda_k=\lambda_k-G_p^{-1}\hat \Gamma \hat d_k$, where $\hat d_k$ is estimated according to (\ref{eq_mod_estimator}),  the estimation error is bounded by
\[ |\hat e_{\infty}|\leq \frac{|\Gamma|  T\mu}{1-|\phi|} \] 
\end{prop}
}

\emph{Proof:}\\

The proof is similar to Proposition \ref{prop1}, but because $\hat \lambda_k=\lambda_k-\hat G_k$ is the input, it leads to 
\begin{align*}
\hat e_{k+1}&=\Gamma d_{k+1}-\hat \Gamma K\hat e_k-\hat\Gamma \hat d_k\\
         &=\Gamma\Delta d_{k+1}+\hat \phi e_k
\end{align*}

\normalsize
As $K=(1-\phi)/\hat \Gamma $, in the equilibrium when $\hat e_{k+1}=\hat e_{k-1}$, $\hat e_{\infty}$ is bounded by
\[|\hat e_{\infty}|=\frac{\Gamma \Delta d_{k+1}}{1-|\phi|}\leq \frac{\Gamma T\mu}{1-|\phi|}
\]
which satisfies the proof. \hfill $\blacksquare$

According to Proposition \ref{prop2}, the z transform of the error $\hat e_k$ is

\[ \hat e(z)=\frac{G(z-1)}{z-\phi}d(z)\]
and the new sensitivity function $\widetilde S_{\epsilon,d}(z)$ can be obtained as follows
\[\mathcal{E}(z)= -\frac{2\eta\mathcal{E}(z)}{(1-z^{-1})z}+\hat e(z)
\]
Dividing by $d(z)$  and factorizing we obtain 
\begin{equation}\label{eq_hat_sed}
\frac{\mathcal{E}(z)}{d(z)}=\hat S_{\epsilon,d}=\frac{G(z-1)^2}{(z-\phi)(z-1+2\eta)}\end{equation}.

\begin{figure}[htbp]
\centering
\includegraphics[width=\columnwidth]{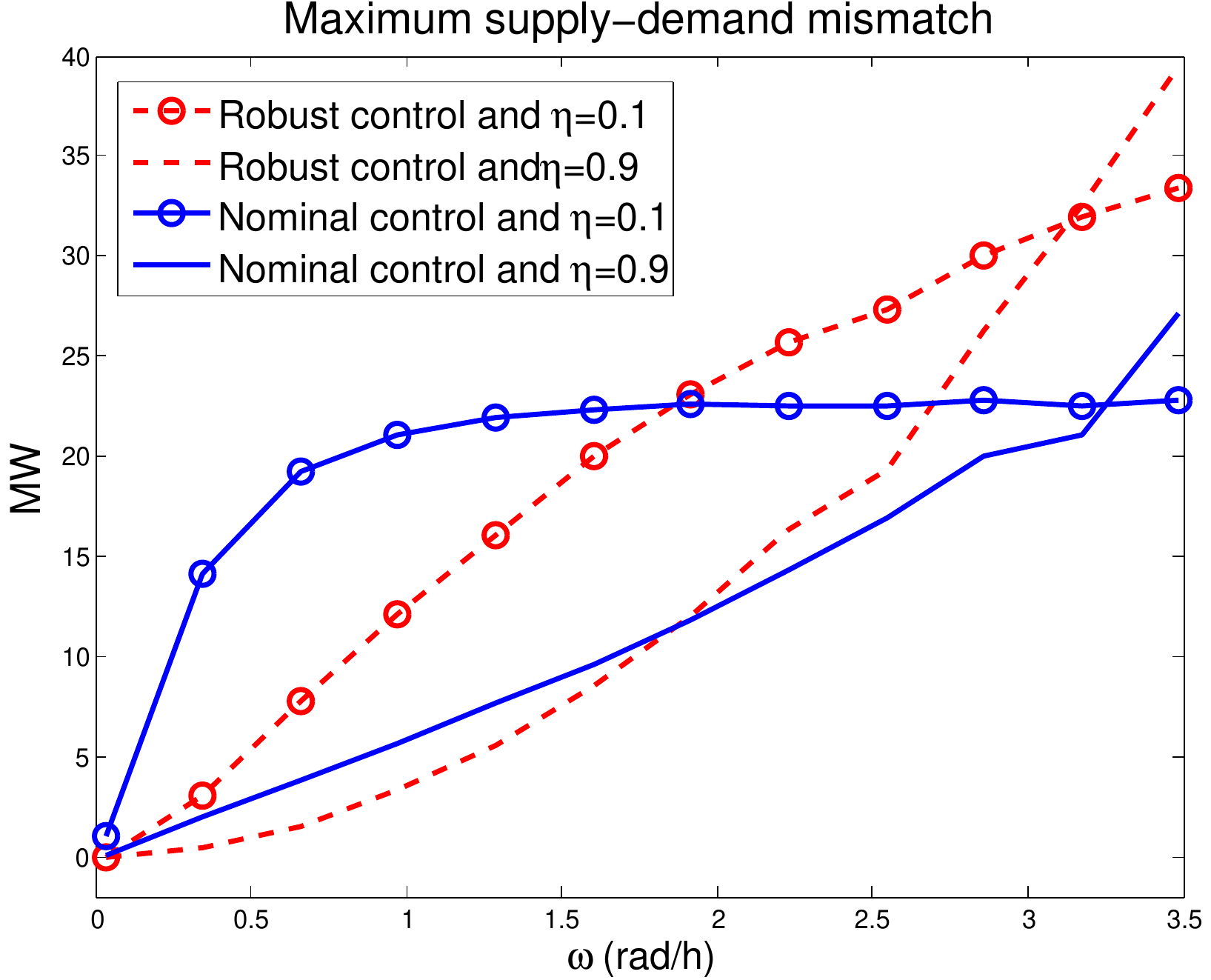}
\caption{Maximum supply-demand mismatch with the nominal controller and with the robust controller. We can see that our robust controller design can attenuate the errors caused by the attack; however, at high frequencies it increases the errors. In a later section we propose the use of low-pass filters to prevent an attacker from using high-frequency attacks.  }
\label{Fig_comparison_robust}
\end{figure}

\begin{figure}[htbp]
\centering
\includegraphics[width=\columnwidth]{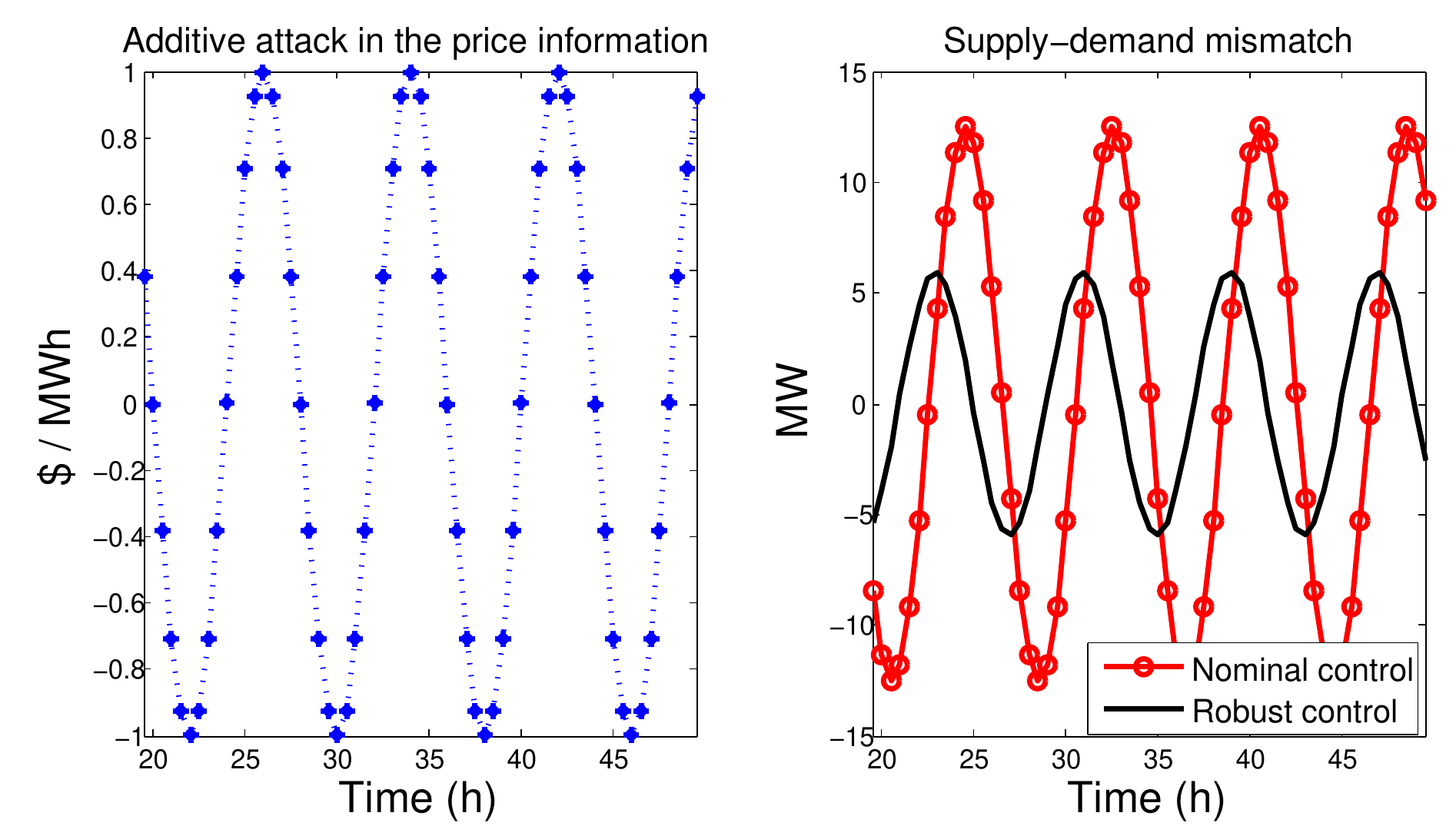}
\caption{ Supply-demand mismatch for an attack of $d_k=\sin(\pi/4kT)$. We can see that our new robust controller attenuates the supply-demand mismatch better than the nominal controller.}
\label{Fig_comparison_robust_time}
\end{figure}

Figures \ref{Fig_comparison_robust} and \ref{Fig_comparison_robust_time} illustrate the maximum supply-demand mismatch with the robust controller and the nominal controller for an attack of the form  $\sin(\omega kT)$.  Note that for frequencies below 1.9, the attack attenuation is better than without the add-on compensator; however, for high frequencies, the inclusion of the compensator increases the impact of the attack.

\begin{figure}[htb]
  \centering
  \includegraphics[width=\columnwidth]{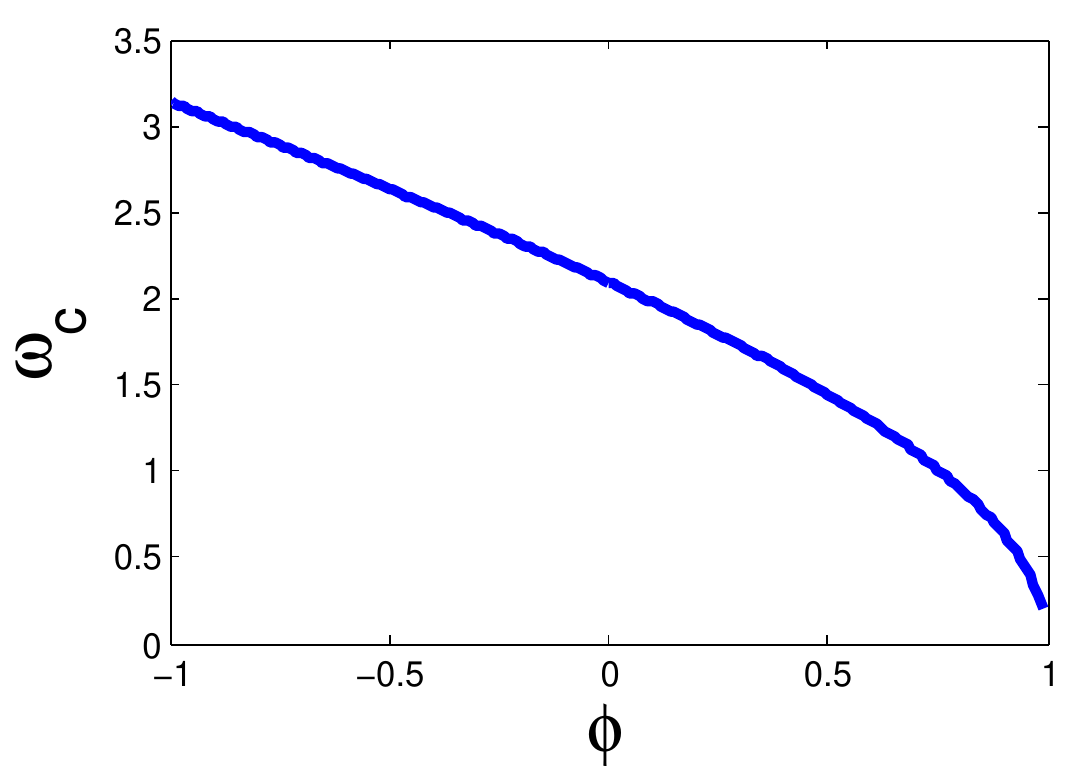}\\
  \caption{Cut-off frequency (for the usefulness of the robust controller) depends on $\phi$.}\label{wc_vs_phi}
\end{figure}

We can obtain the frequency at which the robust controller stops improving the system response under attacks. To do this, we need to find $\omega_c=\omega: |S_{\epsilon,d}(j\omega T)|=|\hat S_{\epsilon,d}(j\omega T)|$. Taking Equation (\ref{eq_sed}) and (\ref{eq_hat_sed}), we have that 
\[|z-1|=|z-\phi|
\]
Replacing $z=e^{j\omega T}$ and solving for $w_c$, we obtain 
\[\omega_c=\frac{1}{T}\arccos \left(\frac{\phi^2-1}{2(\phi-1)}\right)
\]
This relationship is shown in Figure~\ref{wc_vs_phi}. Note that this frequency depends on  $\phi$. $\omega_c$ is larger when $\phi$ approaches $-1$. However, the pole corresponding to $z-\phi$ would approach the unit circle, compromising the exponential stability of the system.

\subsection{Improving Robustness:  Digital Low-Pass Filters in the Smart Meters}

According to Figure~\ref{wc_vs_phi}, the maximum frequency where our proposed robust controller can improve the performance and attenuate the supply-demand error under our attacks is $\omega_c=\pi/(2T) $. The maximum frequency at which an attacker can generate an additive attack is $\omega_{max}=\pi/T$. So, there is a range of frequencies that are amplified by the robust controller. To mitigate this issue we propose the use of a digital low-pass filter in the smart meters, in order to filter  price information with high frequency components.  The same filter has to be implemented by the third party that calculates the price. The cut-off frequency  is given by $\omega_c$. Therefore, for our robust controller to work, we conclude that every frequency greater than $\omega_c$ should be attenuated by the low-pass filter. \\

\begin{figure}[htbp]
\centering
\includegraphics[width=\columnwidth]{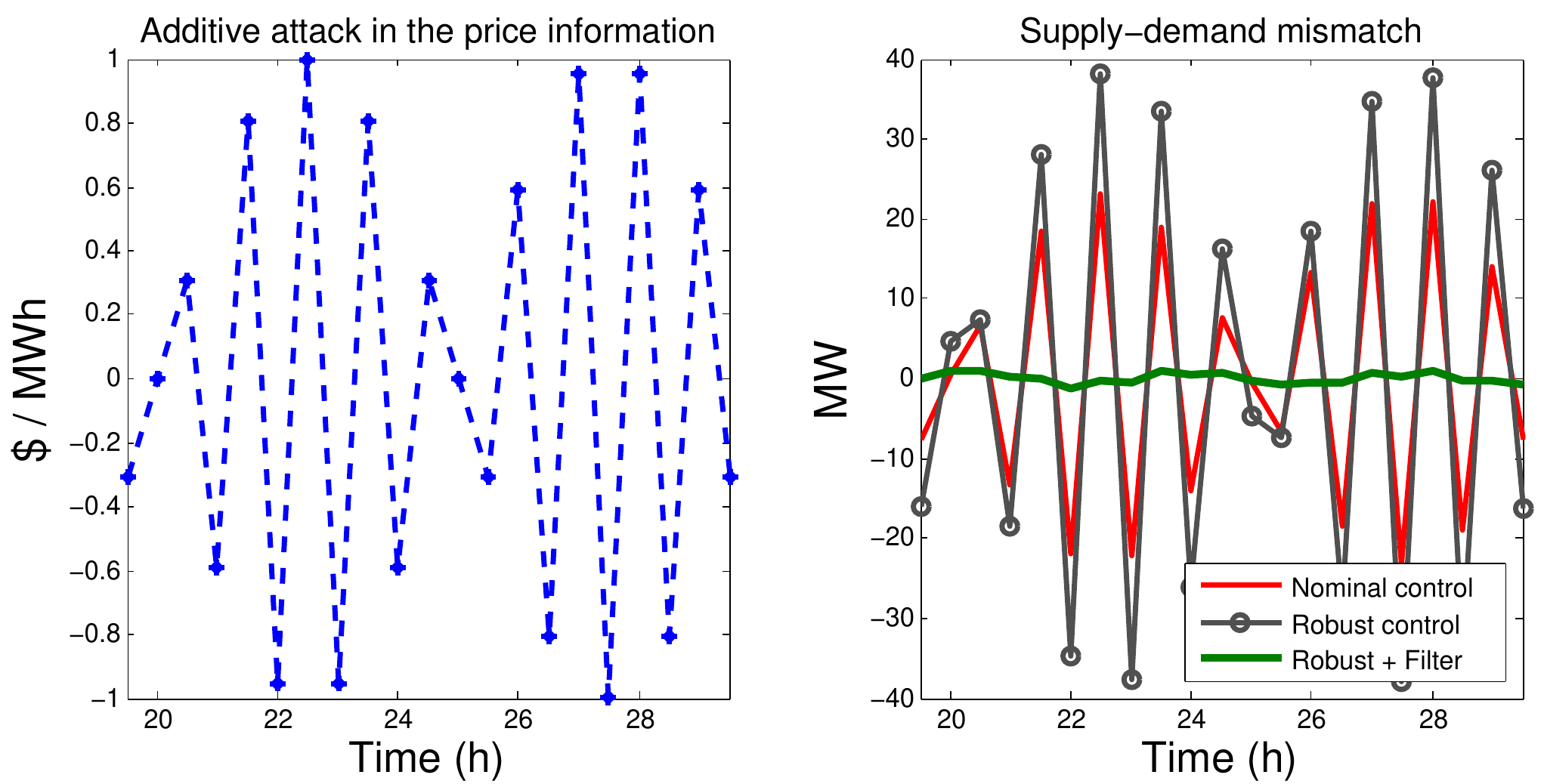}
\caption{ Supply-demand mismatch for $d_k=\sin(2\pi kT)$ using the nominal controller, the robust controller, and the robust+filter control.}
\label{Fig_comparison_robust2}
\end{figure}

We now compare the performance and robustness of the real-time pricing model including a digital IIR low pass filter (Figure~\ref{Fig_comparison_robust2}).  The mathematical analysis for designing the filter is omitted because this topic is out of the scope of this manuscript. The reader only needs to know that there are filters that can eliminate high-frequency components of any signal. 

Admittedly we could also have proposed deploying low-pass filters at the beginning of the paper (before the design of the robust controller), and we could also have seen a significant improvement in minimizing the maximum error that an attacker can create. However, as Figure~\ref{Fig_comparison_robust} shows, the performance of the robust controller for low-frequency signals is still better than the performance of the controller proposed in previous work; therefore with the combination of low-pass filters and robust controllers we seem to have obtained an ideal combination of protection mechanisms.

In summary, the combination of a low-pass filter deployed at all smart meters (or all devices receiving price signals) in addition to a robust controller seems to be the best solution to attenuate any type of attack against our system. We believe this is one of the few instances where a proposed attack-resilient control algorithm does not pose significant negative performance impacts on the system (when the system is not under attack), but we plan to continue evaluating our algorithm in other realistic real-time pricing settings to identify any limitations. 

\section{Detection mechanism}
We have designed a new real time pricing algorithm that not only assures stability, but also minimizes the impact of  attacks.  However, in practice, while we have attenuated the attack, it would still be desirable to know if we are under attack or not, so we can remove compromised devices from our system. 

The ISO calculates a  clearing price each time period, but even in the presence of an attack, the price changes are small (see Figures \ref{fig_example1} and \ref{fig_example2}). However, the state estimator used in our robust controller can give information about the presence of an attacker, by analyzing the statistical behavior of the state estimator over long periods of time.

The detection  mechanism that we propose is based on the accumulation of the rate of change of the estimated signal $\hat d_k$. This is known as the non-parametric CUSUM change detection statistic, and it is defined as:
\begin{eqnarray}\label{eq_detection}
S_0&=&0 \nonumber \\
S_{k+1}&=&(S_k+|\hat G\hat d_k-\hat G\hat d_{k-1} |-\alpha_k)^+
\end{eqnarray}
where $S_k$ is the accumulated impact of the disturbance, and $\alpha_k$ is the rate of change of $S_k$ under normal conditions (without attacks).
The use of the the error $\hat \Gamma \hat d_k$ is due to the fact that the ISO does not have knowledge about $\Gamma$.  
An attack is detected when   $S_k>\delta$. $\delta$ has to be selected such that the number of false alarms is low. As it is based on the rate of change, then high frequency attacks are detected faster.



 
\subsection{Simulations: Detecting attacks}

We assume a populated area with 1 million households, each one receiving information about the price every 30 minutes. To improve the realism of the simulations,  we  assume that  the parameters $D$ and $b_k$ change each time period according to a half-hourly baseline demand profile provided by AEMO from July 21st to 27th, in NSW, Australia.   The baseline load per house is a scaled version of the real whole  NSW region. The parameters of the linear CEO model are $p=31$ and $q=917$  during the simulation time.  

We assume that an attack is launched and modifies the price information of 50\% of the households. The attack is of amplitude $0.1$ \$ /MWh, and a frequency $\omega$. 

The estimation is based on prior information of the baseline load. However, we assume an error in the real-time baseline consumption, such that the ISO calculates the estimation and the robust control based on an approximate load profile, and not the real time consumption. Despite that limitation, the detection algorithm is able to detect an attack when a threshold is achieved.

\begin{figure}[htb]
  \centering
  \includegraphics[width=\columnwidth]{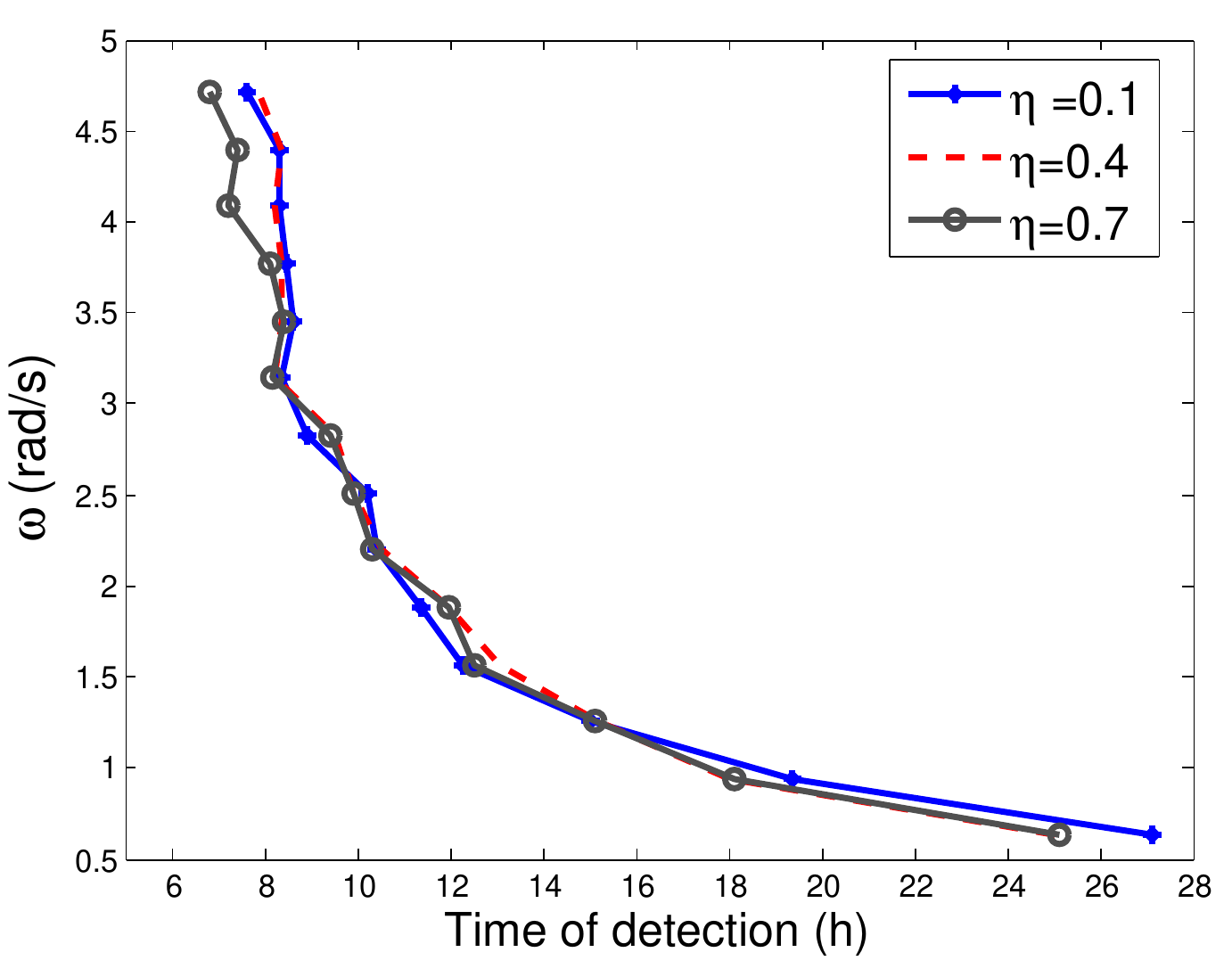}\\
  \caption{Detection for different frequency values of the attack.}\label{fig_time_detection}
\end{figure}

Figure \ref{fig_time_detection} illustrates the time that it takes to detect an attack depending on the frequency of the attack for a threshold $\delta = 10$, which is selected with results without attack in order to avoid false alarms.  Note that for high frequency, the time of detection is low, which is an advantage in order to start a scan in the smart meters and find the victims of the attack.

We can also observe that the detection time does not depend on the $\eta$. 

Our work on detection is preliminary, and in future work we plan to identify the tradeoffs the attacker will face when deciding to launch attacks that maximize the error between power generated and consumed while also maintaining the attack undetected.

\section{Conclusions}

In this work we used the theory from sensitivity analysis to understand how previously proposed attacks could be generalized and evaluated in a formal setting. In particular we showed how to find better attacks than previously proposed, and how to design robust control systems that can mitigate a large number of attacks.



We also found that the design of the price adjustment mechanism is fundamental in the resiliency of the system. In particular, low values of $\eta$ reduce the effect of the attacks on both the prices and sensors.

Another of our contributions was the model to sensor attacks, and how they can have potentially more damaging effects than attacks on the pricing signal. 

We also proposed an attack-resilient controller and several mitigation mechanisms, such as the use of low-pass filters to prevent high-frequency signals, and attack detection mechanisms.  We believe we are one of the few research papers focusing on the important aspect of designing robust control algorithms against false data injection, as much of the previous work tends to focus on state estimation but does not consider the control actions of the system under attack, and how to design an controller that mitigates these attacks. 

Our results show principled ways to use control theory in the design of attack-resilient cyber-physical systems. In general we believe that a well-designed defense-in-depth mechanism for cyber-physical systems will have to leverage not only information security expertise, but control theory to detect, respond, and reconfigure systems that can survive partial compromises. 

Successfully compromising computers and embedded systems participating in controlling the power grid is only the first step to a successful attack.  To have a predictable \emph{physical} modification to the power grid (e.g., strategically manipulating voltages, or loads), the attacker needs to understand how control systems operate. 

Defenders that leverage only information security mechanisms in their protection strategy will have limited success against these sophisticated attackers. To develop a defense-in-depth security strategy, defenders need to incorporate control models of the power grid to understand the vulnerabilities and fragility of the system they are trying to protect (e.g., not all compromised devices can drive a system to an unsafe state), and to design attack-resilient control algorithms that can survive a partial compromise of the system.  Our work shows a direction of how to pursue this goal further and in general we hope these formalisms can help mitigate attacks not only against the power-grid but against other cyber-physical systems.

One interesting area of future research that we did not address in this paper are the possible attack strategies that can be achieved by combining attacks to both: sensors and control signals. All our models assumed the attacker compromised either the price signals, or the sensor signals, but not both. It is clear that if the attacker controls all control signals and all sensor signals then there is nothing we can do, but if the attacker has partial compromise of controllers and sensors, then the defender might still be able to design a robust algorithm that attenuates the attacks. We plan to look into this area in future work. 


\bibliographystyle{plain}
\bibliography{Bibliography_Jairo,alvaro,alvaro2,ccs,demand,lagrange,literature,references_b,refs,refsA,smartgrid,traffic}

\end{document}